\documentclass[letterpaper,11pt]{article}
\usepackage{jheppub}
\usepackage{tikz-cd}
\usepackage[dvipsnames]{xcolor}
\usetikzlibrary{calc, shapes.misc}
\usepackage{orcidlink}
\tikzset{
			base/.style={baseline=-0.6ex, scale=1},
			dot/.style={circle, fill=black, inner sep=1.5pt, outer sep=0pt},
			cross/.style={font=\bfseries\scriptsize, inner sep=0pt},
			tube/.style={draw, thick, rounded corners=4.5pt, line width=0.8pt},
			chain/.style={thick}
		}

\def\oh{\mathcal{O}}
\def\nh{\mathcal{N}}
\newcommand{\twbr}[1]{\langle #1 \rangle}

\title{ Landau Analysis of One-Cycle Negative Geometries}

\author[a]{\small Shruti Paranjape\,\orcidlink{0000-0003-2613-718X},}
\emailAdd{shruti\_paranjape@brown.edu}

\author[a]{\small Marcos Skowronek\,\orcidlink{0000-0003-2569-1234},}
\emailAdd{marcos\_skowronek\_santos@brown.edu}

\author[a, b]{\small Marcus Spradlin\,\orcidlink{0009-0005-6084-2466},}
\emailAdd{marcus\_spradlin@brown.edu}

\author[a]{\small Anastasia Volovich\,\orcidlink{0009-0008-2506-3207}}
\emailAdd{anastasia\_volovich@brown.edu}

\author[a]{\small and He-Chen Weng\,\orcidlink{0000-0003-3781-6153}}
\emailAdd{he-chen\_weng@brown.edu}

\affiliation[a]{\footnotesize Department of Physics,
    Brown University,
    Providence,
    RI 02912,
    USA
}

\affiliation[b]{\footnotesize Brown Center for Theoretical Physics and Innovation,
    Brown University,
    Providence,
    RI 02912,
    USA
}

\abstract{We use geometric Landau analysis to determine the singularity structure of four-point, one-cycle negative geometries in $\nh=4$ super-Yang-Mills theory, which represent certain contributions to the logarithm of the four-point amplitude or equivalently the normalized quadrangular Wilson loop with a Lagrangian insertion. By analyzing the relevant Landau diagrams recursively, we prove that this quantity has singularities only at $z=-1,0$ and $\infty$ to all loop orders. This represents a first step towards obtaining a non-perturbative resummation for this quantity at next-to-leading order in the expansion over cycles.}

\begin{document}
\maketitle
\flushbottom

\section{Introduction}\label{sec: intro}

Scattering amplitudes in planar $\mathcal{N}=4$ supersymmetric Yang-Mills (SYM) theory have many interesting mathematical constructions and properties (see \cite{Arkani-Hamed:2022rwr} for a review). On one hand, at weak coupling, the symbol bootstrap program~\cite{Goncharov:2010jf,
Dixon:2011pw,Dixon:2011nj,Golden:2013xva,Dixon:2013eka,Dixon:2014voa,Dixon:2014iba,Drummond:2014ffa,Dixon:2015iva,Caron-Huot:2016owq,Dixon:2016apl,Dixon:2016nkn,Caron-Huot:2019vjl, Dixon:2020cnr, Caron-Huot:2020bkp} has enabled the six-point amplitude to be computed to eight loops~\cite{Dixon:2023kop} and the seven-point amplitude to five loops~\cite{He:2025tyv}. Meanwhile, the amplituhedron~\cite{Arkani-Hamed:2013jha,Arkani-Hamed:2013kca,Arkani-Hamed:2017vfh, Arkani-Hamed:2018rsk} encodes a geometric construction of integrands to all loop orders. On the other hand, integrability of the theory in the planar limit has allowed for the extraction of numerous strong- and finite-coupling results via e.g.~the BES equation~\cite{Beisert:2006ez}, surfaces in AdS space~\cite{Alday:2007hr,Alday:2010vh}, the pentagon operator product~\cite{Alday:2010ku,Basso:2013vsa,Basso:2013aha,Basso:2014koa, Basso:2014nra,Basso:2014pla}
and origin kinematics expansions~\cite{Basso:2020xts, Basso:2022ruw}.

In \cite{Arkani-Hamed:2021iya}, Arkani-Hamed, Henn and Trnka connected these regimes by resumming subsets of perturbative contributions to the logarithm of the four-point amplitude, which is dual to a null polygonal Wilson loop with one Lagrangian insertion. This function of a single variable is typically denoted $F(z)$ in the literature. 
The construction proceeds via defining \emph{negative geometries}, a modified version of the loop amplituhedron in which pairs of loop momenta are not all required to be mutually positive, but satisfy negativity conditions encoded in certain graphs where each node represents one loop. The complexity of a term contributing to the expansion of $F(z)$ is directly related to the number of internal cycles in the corresponding graph. By resumming the terms associated to tree graphs with arbitrarily many nodes, i.e. to all loop orders, they obtained non-perturbative contributions to $F(z)$ and hence to the cusp anomalous dimension, supplementing previous calculations \cite{Bern:2006ew, Benna:2006nd,Basso:2007wd}. They found that these ``tree-graph negative geometry'' contributions to the cusp anomalous dimension have the correct strong-coupling scaling expected from AdS/CFT (but with a different coefficient) and agree with its perturbative expansion at weak coupling.

Since then, there has been great effort in characterizing more complicated terms in the expansion of $F(z)$ over negative geometries. In \cite{Brown:2023mqi}, an expression for the contribution of negative geometries containing one closed cycle to the integrand of $F(z)$ was obtained at arbitrary loop order. Explicit results for the integrated one-cycle negative geometries at two and three loops \cite{Dixon:2026} show that they can be written purely in terms of harmonic polylogarithm functions, and so they only have branch point singularities at a few particular locations: $z\in \{-1,0,\infty\}$. However, it is still unknown what the exact form of the integrated result is for higher loops, and in particular if the singularity structure remains as simple as in the lower-loop cases.

In this paper we provide an affirmative answer to the last question by proving that the locus of singularities for the one-cycle negative geometry at any loop order is restricted to $z \in \{-1,0,\infty\}$. In order to do that, we utilize the tools of \emph{geometric Landau analysis}, first introduced in \cite{Dennen:2015bet,Dennen:2016mdk,Prlina:2017azl,Prlina:2017tvx,Lippstreu:2022bib,Lippstreu:2023oio} to study planar amplitudes in SYM. This approach combines solving the usual Landau equations~\cite{Landau:1959fi} to find potential singularities, together with imposing the constraints of the geometry in order to determine which ones are physical or spurious. This approach has been recently used~\cite{Chicherin:2025cua} as input to help determine the symbol~\cite{Goncharov:2010jf} of two- and three-loop ladder negative geometries at higher multiplicity. Several recent works have explored various algorithms for and applications of Landau analysis, including~\cite{Correia:2020xtr,Hannesdottir:2021kpd,Dlapa:2023cvx,Fevola:2023fzn,Hannesdottir:2024hke,He:2024fij,Correia:2025wtb,Hollering:2026gjz,Hollering:2026rlv}.

The paper is organized as follows. In Section~\ref{sec: review} we provide a brief description of the fundamental objects we work with and establish some key notation. In Section~\ref{sec: examples} we illustrate the geometric Landau analysis by solving some explicit examples at low loop order, while in Section~\ref{sec: general claims} we systematically study different classes of topologies in Landau diagrams for arbitrary number of loops. Our main result is a recursive proof that the complete list of singularities is $z \in \{-1,0,\infty\}$. Finally, we conclude in Section~\ref{sec: outlook} with an outlook.

\section{Key concepts and notation}\label{sec: review}

In this section, we briefly review the fundamental objects and tools that we make use of in this work: momentum twistor variables in Section~\ref{sec: momentum twistor}, negative geometries in Section~\ref{sec: amplituhedron}, and the use of geometric Landau analysis to diagnose physical and spurious singularities in Section~\ref{sec: landau}.

\subsection{Momentum twistors}\label{sec: momentum twistor}

Momentum twistors~\cite{Hodges:2009hk} are based on the equivalence between null rays in complexified Minkowski space and points in twistor space $\mathbb{P}^3$ or, equivalently, lines in twistor space and points in Minkowski space. We use the notation $Z_i, Z_j,\ldots$ to denote points in $\mathbb{P}^3$. We can further parameterize each point with  four-component homogeneous coordinates $Z_i=(Z_i^1,Z_i^2,Z_i^3,Z_i^4)$, up to the identification $Z_i\sim t Z_i$ for any non-zero complex number $t$. We also denote the bi-twistor $\epsilon_{ABCD} Z_i^C Z_j^D$ with the shorthand notation $(ij)$. Geometrically, $(ij)$ corresponds to the line in $\mathbb{P}^3$ containing $Z_i$ and $Z_j$. Similarly, we denote $\epsilon_{ABCD} Z_i^B Z_j^C Z_k^D$ by $(ijk)$, which corresponds to the plane containing $Z_i$, $Z_j$ and $Z_k$ in $\mathbb{P}^3$.

The natural $SL(4,\mathbb{C})$-invariant is expressed as the twistor bracket
\begin{equation}
    \langle ijkl\rangle = \epsilon_{ABCD}Z_i^A Z_j^B Z_k^C Z_l^D\,.
\end{equation}
We are often interested in the geometric interpretation of the locus where a four-bracket vanishes. For instance, $\langle ijkl\rangle = 0$ means that two lines $(ij)$ and $(kl)$ intersect. Equivalently, the two lines $(ik)$ and $(jl)$ intersect, the point $Z_i$ lies on the plane $(jkl)$, and so on.

Finally, we present a couple of identities that will prove useful in later sections. Firstly, because five different twistors have to be linearly dependent, we have
\begin{equation}
    \twbr{ijk a}\twbr{bcde} + \twbr{ijk b}\twbr{cdea} + \twbr{ijkc}\twbr{deab} + \twbr{ijkd}\twbr{eabc} + \twbr{ijke}\twbr{abcd}=0\,.
\end{equation}
This is simply the four-dimensional version of the Schouten identity for spinor products. In particular, for any line $AB$, we have
\begin{equation}
    \twbr{AB12}\twbr{AB34} + \twbr{AB23}\twbr{AB14} - \twbr{AB13}\twbr{AB24} = 0\,,
\end{equation}
which we will refer to simply as the Schouten identity. Next, the intersection of two planes $(abc) \cap (ijk)$ gives a line that satisfies the relation
\begin{equation}
    \twbr{x\,y\,(abc)\cap(ijk)} = \twbr{x\,a\,b\,c}\twbr{y\,i\,j\,k}-\twbr{y\,a\,b\,c}\twbr{x\,i\,j\,k}\,.
\end{equation}
A special case is $k=c$ when the lines $(ab)$ and $(ij)$ intersect, in which case
\begin{equation}\label{eq:plane int id}
    \twbr{x\,y\,(abc)\cap(ijc)} = \twbr{x\,y\,c\,(ab)\cap(ij)}= \twbr{x\,a\,b\,c}\twbr{y\,i\,j\,c}-\twbr{y\,a\,b\,c}\twbr{x\,i\,j\,c}\,.
\end{equation}

\subsection{Amplituhedron and negative geometries}\label{sec: amplituhedron}

The integrand for an $L$-loop amplitude in planar $\nh=4$ SYM theory can be expressed as the canonical differential form on a geometry called the \emph{amplituhedron} \cite{Arkani-Hamed:2013jha, Arkani-Hamed:2013kca,Arkani-Hamed:2017vfh}. At four points, it is defined as the configuration space of $L$ ``loop lines'' $AB_i$ in momentum twistor space, together with external kinematic data $\{Z_1,\,Z_2,\,Z_3,\,Z_4\}$ that satisfy the positivity condition $\twbr{1234}>0$. In addition, the loop lines must obey the inequalities
\begin{equation}\label{eq: amplituhedron ext positivity}
    \twbr{AB_i 12},\,\twbr{AB_i 23},\, \twbr{AB_i 34},\, \twbr{AB_i 14}>0\,,\quad \twbr{AB_i 13},\,\twbr{AB_i 24}<0\,,\quad i=1,\ldots,L\,,
\end{equation}
as well as the multi-loop positivity conditions
\begin{equation}\label{eq: amplituhedron int positivity}
    \twbr{AB_i AB_j} >0\,,\quad \forall\ i\neq j\,.
\end{equation}
The first inequality in~(\ref{eq: amplituhedron ext positivity}) and the one in~(\ref{eq: amplituhedron int positivity}) become equalities if we include the boundaries of the amplituhedron, which will often be useful. The integrand for the four-point $L$-loop amplitude is the unique canonical form $\Omega_L$ with logarithmic singularities on the boundaries of this space. In general, it can be written as
\begin{equation}
    \Omega_L = \frac{d\mu_L \, \nh_L}{\left(\prod_i D_i\right)\left( \prod_{i,j} \twbr{AB_i AB_j}\right)}\,,\quad D_i\equiv \twbr{AB_i 12}\twbr{AB_i 23}\twbr{AB_i 34} \twbr{AB_i 14}\,,
\end{equation}
where the loop integration measure is given by
\begin{equation}
    d\mu_L = \prod_{i=1}^L \twbr{AB_i d^2A}\twbr{AB_i d^2B}\,.
\end{equation}
The form of the numerator $\nh_L$ can be fixed, in principle, by demanding that the integrand only has logarithmic singularities on the boundaries of the amplituhedron \cite{Arkani-Hamed:2018rsk}, which gets increasingly more complicated to do at higher loops. For our purposes it will suffice to use the defining geometric constraints \eqref{eq: amplituhedron ext positivity}--\eqref{eq: amplituhedron int positivity}, which will allow us to rule out some putative Landau singularities as unphysical~\cite{Dennen:2016mdk}.

Since the IR divergences at $L$-loop order (and in fact the full kinematic dependence, up to constants \cite{Bern:2005iz}) in SYM exponentiate, it is convenient to work with the logarithm of the amplitude, which only has a mild $1/\epsilon^2$ divergence,
\begin{equation}
    \log M_L = \frac{\gamma_{\text{cusp}}}{\epsilon^2} + \oh\left(\frac{1}{\epsilon}\right),
\end{equation}
where $\gamma_\text{cusp}$ is the cusp anomalous dimension. In fact, if at 
$(L{+}1)$-th loop order we keep the last loop line $AB_{L+1}\equiv AB_0$ unintegrated, then the resulting object $F^{(L)}$ will be IR finite~\cite{Arkani-Hamed:2021iya}. By the amplitude/Wilson loop duality \cite{Alday:2007hr,Alday:2011ga,Drummond:2007cf,Brandhuber:2007yx,Drummond:2007aua,Engelund:2011fg,Engelund:2012re}, it is equal to the $L$-loop order term in the perturbative expansion of the normalized polygonal Wilson loop with a Lagrangian insertion,
\begin{equation}
    F(z) = \pi^2\frac{x_{10}^2x_{20}^2x_{30}^2x_{40}^2}{x_{13}^2x_{24}^2}\frac{\langle W_4(x_1,\,x_2,\,x_3,\,x_4)\mathcal{L}(x_0)\rangle}{\langle W_4(x_1,\,x_2,\,x_3,\,x_4)\rangle}\,.
\end{equation}
Because of dual conformal symmetry, this quantity only depends on the single variable
\begin{equation}\label{eq: z variable}
    z=\frac{\twbr{AB_012}\twbr{AB_034}}{\twbr{AB_023}\twbr{AB_014}} = -1 + \frac{\twbr{AB_013}\twbr{AB_024}}{\twbr{AB_023}\twbr{AB_014}}\,.
\end{equation}
In \cite{Arkani-Hamed:2021iya}, the authors showed that the $L$-loop contribution to $F(z)$ can be written as an expansion over so-called \emph{connected negative geometry graphs}:
\begin{equation}
    F(z) = \sum_{\text{connected graphs}\ G} (-1)^{E(G)}\hspace{6pt} \begin{tikzpicture}[baseline={([xshift=3pt,yshift=-0.5ex]current bounding box.center)}, scale=1]
  % Define the style for the black nodes
  \tikzset{dot/.style={circle, fill=black, inner sep=2.5pt},cross/.style={path picture={\draw[thick,black](path picture bounding box.south east) -- (path picture bounding box.north west) (path picture bounding box.south west) -- (path picture bounding box.north east);
  }}}
  
  % Place the 5 vertices of the pentagon
  \node[dot] (n1) at (90:1) {};
  \node[dot] (n2) at (18:1) {};
  \node[dot] (n3) at (-54:1) {};
  \node[draw,thick,circle,cross,minimum width=8pt] (n4) at (-126:1) {};
  \node[dot] (n5) at (162:1) {};
  
  % Draw the perimeter edges and the internal diagonals
  % Using red!50!black to match the dark red color in your image
  \begin{scope}[very thick, Maroon]
    % Outer cycle
    \draw (n1) -- (n2) -- (n5) -- (n4) -- (n3) -- (n5);
  \end{scope}
\end{tikzpicture}
\end{equation}
where each node represents a loop line $AB_i$, and we have denoted the unintegrated loop $AB_0$ with a cross. In this expansion the multi-loop positivity conditions defining the amplituhedron are replaced by either no condition on $\langle AB_i AB_j\rangle$, if nodes $i$ and $j$ are not connected by an edge, or a negativity condition for connected nodes:
\begin{equation}\label{eq: neg geometry expansion}
    \twbr{AB_i AB_j}<0 \quad \text{for every} \quad \begin{tikzpicture}[baseline={([xshift=3pt,yshift=5pt]current bounding box.center)}, scale=1]
  % Define the style for the black nodes
  \tikzset{dot/.style={circle, fill=black, inner sep=2.5pt},cross/.style={path picture={\draw[thick,black](path picture bounding box.south east) -- (path picture bounding box.north west) (path picture bounding box.south west) -- (path picture bounding box.north east);
  }}}
  
  % Place the 5 vertices of the pentagon
  \node[dot] (n1) at (180:1) {};
  \node[dot] (n2) at (0:1) {};
  \node[below=4pt] at (n1) {$AB_i$};
  \node[below=4pt] at (n2) {$AB_j$};
  
  % Draw the perimeter edges and the internal diagonals
  % Using red!50!black to match the dark red color in your image
  \begin{scope}[very thick, Maroon]
    % Outer cycle
    \draw (n1) -- (n2);
  \end{scope}
\end{tikzpicture}.
\end{equation}
Importantly, the IR finiteness of these negative geometries is encoded in the fact that the negativity conditions prevent collinear configurations of the loop lines, which we will use to discard spurious singularities.

In this paper we consider one particular class of negative geometries: the $L$-loop four-point one-cycle, whose integrand is given by
\begin{equation}
  \begin{aligned}
    % TikZ figure of the polygon
    \begin{tikzpicture}[baseline=(O.base), line width=0.8pt]
        % Center node for baseline alignment
        \node (O) at (0, 0) {};
        
        % Points for vertices
        \coordinate (P0) at (0,-1.2);
        \coordinate (P1) at (1,-0.7);
        \coordinate (P2) at (1.3,0.3);
        \coordinate (P3) at (0.8,1.3);
        \coordinate (PL-1) at (-1.3, 0.3);
        \coordinate (PL) at (-1, -0.7);
        \coordinate (D1) at (-0.5, 0.9);
        \coordinate (D2) at (-0.35, 1);
        \coordinate (D3) at (-0.2, 1.1);

        % Draw edges
        \draw[very thick, Maroon] (PL-1) -- (PL) -- (P0) -- (P1) -- (P2) -- (P3);
        
        % Draw vertices as small black dots
        \fill (P1) circle (3pt);
        \fill (P2) circle (3pt);
        \fill (P3) circle (3pt);
        \fill (PL-1) circle (3pt);
        \fill (PL) circle (3pt);
        \fill (D1) circle (1pt);
        \fill (D2) circle (1pt);
        \fill (D3) circle (1pt);
        
        % Draw special node at P0 with \otimes symbol style
        \draw[fill=white] (P0) circle (4pt);
        \draw ($(P0)-(2.12pt,2.12pt)$) -- ($(P0)+(2.12pt,2.12pt)$);
        \draw ($(P0)-(2.12pt,-2.12pt)$) -- ($(P0)+(2.12pt,-2.12pt)$);
        
        % Text labels
        \node[anchor=north] at ($(P0)-(0,0.1)$) { $AB_0$};
        \node[anchor=north west] at ($(P1)-(0,0.1)$) { $AB_1$};
        \node[anchor=west] at ($(P2)+(0.1,0)$) { $AB_2$};
        \node[anchor=south west] at ($(P3)+(0.1,0)$) { $AB_3$};
        \node[anchor=south east] at ($(PL-1)-(0.1,0)$) { $AB_{L-1}$};
        \node[anchor=east] at ($(PL)-(0.1,0)$) { $AB_L$};
        
    \end{tikzpicture}
    % Spacing and similarity symbol
    \; \sim \;
    % Equation aligned with the middle of the figure
    \prod_{\ell=0}^L \left( \frac{1}{\langle AB_\ell AB_{\ell+1} \rangle}\prod_{i=1}^4 \frac{1}{\langle AB_\ell\, i\, i{+}1 \rangle} \right)
\end{aligned}
  \label{eq:onecycle}
\end{equation}
where $\sim$ indicates that we have omitted the numerator factor (which was calculated in \cite{Brown:2023mqi}) and have only shown the denominator, which is a product of $5(L+1)$ propagators.  The purpose of this paper is to show that, at any loop order, the integral of this geometry can have branch point singularities only when the variable~(\ref{eq: z variable}) takes values in $z\in\{-1,0,\infty\}$.

\subsection{Landau analysis}\label{sec: landau}

\begin{figure}
    \centering
    \begin{tikzpicture}[scale=1.25]
		\draw[thick](0.5,0.5)--(0.25,0.25);
        \node at (0.15,0.15) {2};
		\draw[thick](0.25,1.75)--(0.5,1.5);
        \node at (0.15,1.85) {3};
		\draw[thick](0.5,0.5)--(1.5,0.5);
		\draw[thick](0.5,0.5)--(0.5,1.5);
		\draw[thick](0.5,1.5)--(1.5,1.5);
		\draw[thick](1.5,0.5)--(1.5,1.5);
		\draw[thick](1.5,0.5)--(2.5,0);
		\draw[thick](1.5,1.5)--(2.5,2);
		\draw[thick](3.25,1)--(2.5,0);
		\draw[thick](3.25,1)--(2.5,2);
		\draw[thick](3.25,1) to[out=-130,in=170] (4,0);
		\draw[thick](3.25,1) to[out=0,in=70] (4,0);
		\draw[thick](2.5,0)--(4,0);
		\draw[thick](2.5,0)--(2.25,-0.25);
		\draw[thick](2.5,0)--(2.75,-0.25);
        \node at (2.15,-0.35) {1};
        \node at (2.85,-0.35) {2};
		\draw[thick](2.5,2)--(2.75,2.25);
        \node at (2.85,2.35) {2};
		\draw[thick](2.5,2)--(2.25,2.25);
        \node at (2.15,2.35) {4};
		\draw[thick](4,0)--(4,-0.5);
        \node at (4,-0.7) {1};
		\draw[thick](4,0)--(4.75,0);
		\draw[thick](3.25,1)--(4.75,1);
		\draw[thick](4.75,0)--(4.75,1);
		\draw[thick](5,1.25)--(4.75,1);
        \node at (5.1,1.35) {3};
		\draw[thick](4.75,0)--(4.9,-0.3);
		\draw[thick](4.75,0)--(5.05,-0.15);
        \node at (5,-0.4) {4};
        \node at (5.15,-0.25) {3};
		\node at (1,1) {\small$AB_1$};
		\node at (0,1) {\small$AB_0$};
		\node at (2.25,1) {\small$AB_2$};
		\node at (3.1,0.2) {\small$AB_3$};
		\node at (3.65,0.5) {\small$AB_4$};
		\node at (4.4,0.5) {\small$AB_5$};
		\node at (5.25,0.5) {\large$\cdots$};
		\draw[thick] (5.75,0)--(5.75,1);
		\draw[thick](6.5,-0.35)--(5.75,0);
		\draw[thick](6.5,1.35)--(5.75,1);
		\draw[thick](6.5,-0.35)--(7.25,0);
		\draw[thick](6.5,1.35)--(7.25,1);
		\draw[thick](7.25,0)--(7.25,1);
		\draw[thick](6.5,-0.35)--(6.5,-0.7);
        \node at (6.5,-0.9) {4};
		\draw[thick](6.5,1.35)--(6.75,1.6);
        \node at (6.85,1.7) {2};
		\draw[thick](6.5,1.35)--(6.25,1.6);
        \node at (6.15,1.7) {1};
		\draw[thick](7.25,0)--(8.25,0);
		\draw[thick](8.25,0)--(8.25,1);
		\draw[thick](8.25,1)--(7.25,1);
		\draw[thick](7.25,0)--(7.4,-0.3);
        \node at (7.5,-0.4) {1};
		\draw[thick](7.25,0)--(7.1,-0.3);
        \node at (7,-0.4) {3};
		\draw[thick](8.25,0)--(8.5,-0.25);
        \node at (8.6,-0.35) {2};
		\draw[thick](8.25,1)--(8.5,1.25);
        \node at (8.6,1.35) {3};
		\node at (6.5,0.5) {\small$AB_{L-1}$};
		\node at (7.75,0.5) {\small$AB_{L}$};
		\node at (8.75,0.5) {\small$AB_0$};
	\end{tikzpicture}
    \caption{Representative example of a Landau diagram for the $L$-loop, one-cycle negative geometry~(\ref{eq:onecycle}).}
    \label{fig: example topology}
\end{figure}
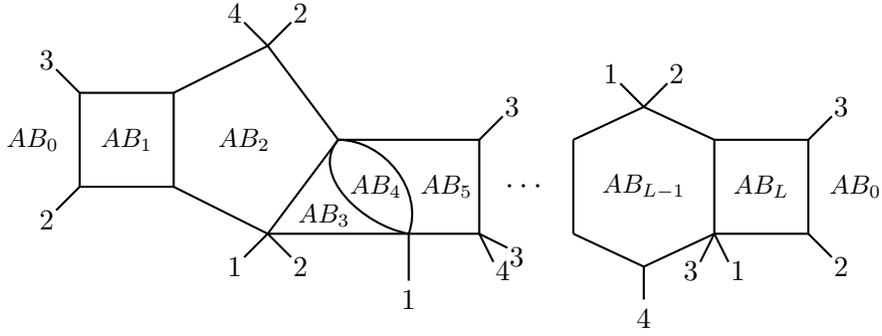

The existence of potential singularities in integrated negative geometries can be diagnosed via standard Landau analysis \cite{Landau:1959fi,Eden:1966dnq}. Potential singularities can be systematically enumerated by considering all subsets of the propagators in which each loop line $AB_\ell$ appears at least twice (this applies for $1 \le \ell \le L$; we exclude the four propagators $\twbr{AB_0\,i\,i{+}1}$ which do not participate in the Landau analysis since $AB_0$ is unintegrated). Each such subset can be graphically represented by a Landau diagram (an example of which is shown in Figure~\ref{fig: example topology}) in which each propagator in the subset is represented by an edge between regions corresponding to the loop lines or external momenta, and each closed region is a bubble, triangle, box, pentagon or hexagon (which is the maximum for~(\ref{eq:onecycle})) depending on how many times the loop line appears in the subset.

Given a Landau diagram of a certain topology, we can think of it as a Feynman diagram and assign a loop momentum $q_e^\mu$ and a Schwinger parameter $\alpha_e$ to each (directed) edge. The location of potential singularities can be found by determining when the following equations admit solutions:
\begin{equation}
\begin{aligned}
    \alpha_e q_e^2=0&, \qquad \forall\, \text{ edges}\, e\quad \text{(cut conditions)},\\
    \sum_{e\in \text{loop}} \alpha_e\, q_e^\mu =0&,\qquad \forall\, \text{ closed loops}\quad \text{(pinch conditions)}.
\end{aligned}
\end{equation}
For solutions with real momenta in Minkowski space and positive values of the Schwinger parameters, these can be interpreted as the conditions for a real spacetime process where all internal particles are on-shell and moving forward in time \cite{Coleman:1965xm}. However, for our purpose of determining all possible singularities of the maximal analytic continuation of the integral, we must allow all quantities to be complex.

Generally the Landau equations only admit solutions on codimension-1 surfaces in the space of external kinematics, on which integrated amplitudes may have branch points.  In our application, the ``external kinematics'' consists of the momentum twistors $Z_1,\ldots,Z_4$ as well as the special, unintegrated loop line $AB_0$. Since we aim to identify potential branch points in the variable $z$ defined in~(\ref{eq: z variable}), we are uninterested in solutions that exist only on surfaces of codimension-2 or higher, or solutions that exist for generic $z$; the latter include solutions that exist only for non-generic $Z_i$ such as when $\langle 1234 \rangle = 0$.   Furthermore, as emphasized in~\cite{Prlina:2017azl}, we are only interested in solutions for which the cut conditions are satisfied for generic external kinematics--it must be the pinch conditions that effectively impose the constraint--since only these can potentially give rise to a branch point. Finally, we also ignore so-called \emph{second-type singularities} \cite{Cutkosky:1960sp}, which arise at infinite values of the loop momenta and thus should be absent in planar SYM when using a regulator that preserves dual conformal symmetry \cite{Dennen:2015bet}.

Each of the cut equations can be solved either by putting the corresponding propagator on-shell $q_e^2=0$, or by setting $\alpha_e=0$.  However, any solution in which one or more of the $\alpha_e$ is zero, which we call a subleading singularity, is also a solution of the Landau equations associated with the subtopology where the corresponding edges are pinched (or equivalently, the propagators are omitted from the corresponding subset).  Our exposition will be streamlined by considering, for each graph, only its leading Landau singularities, for which all of the $\alpha_e$ are nonzero.  This differs from the bookkeeping used for example in~\cite{Dennen:2016mdk,Prlina:2017azl,Prlina:2017tvx} but is similar to the approach used in~\cite{He:2024fij}. One important subtlety in this approach is that it is well-known that for some topologies there are codimension-1 solutions of the Landau equations for which both $q_e^2 = 0$ and $\alpha_e = 0$, but the corresponding subtopology solution is codimension-0 (i.e., generic); such singularities would be overlooked by the algorithm we have described.  However, for the application to~(\ref{eq:onecycle}) we will be able to recursively check that nothing of this type is missed.

As emphasized above, solutions to the Landau equations only indicate the locations of \emph{potential} singularities. The existence or absence of a potential singularity is determined by the behavior of the numerator factor, which is difficult to compute in general.  Instead, once we have worked out the solutions of the Landau equations for a given graph, we can use the geometric criterion formulated in~\cite{Dennen:2016mdk,Prlina:2017azl,Prlina:2017tvx} to determine which ones are physical and which are spurious.  As explained in those references, the key idea is that if the locus of on-shell loop momenta associated with a given solution of the cut conditions lies inside the negative geometry, then the singularity is physical. Otherwise, the monodromy around the putative branch point would vanish, and the singularity is spurious. In other words, a solution to the cut conditions (but not necessarily the pinch conditions) must correspond to an actual boundary of the negative geometry.

\section{Explicit examples}\label{sec: examples}

As a warm-up, we illustrate our use of Landau analysis by explicitly computing the singularities of several diagrams relevant for one- and two-loop negative geometries.

\subsection{One-loop negative geometry}\label{sec: one-loop analysis}

At one loop there is a single negative geometry, and the associated integrand contains the following poles (using here the notation $AB = AB_0$ and $CD = AB_1$ compared to~(\ref{eq:onecycle})):
\begin{equation}
    \begin{tikzpicture}[baseline={([yshift=0.75ex]current bounding box.center)}, 
    thick, scale=0.9,
    cross/.style={path picture={
      \draw[black] (path picture bounding box.south east) -- (path picture bounding box.north west) (path picture bounding box.south west) -- (path picture bounding box.north east);
    }}]
    \coordinate (AB) at (0,0);
    \coordinate (AB1) at (6pt,0pt);
    \coordinate (CD) at (1.5,0);
    % Draw the square
    \draw[very thick, Maroon] (AB1) -- (CD);
    % Draw the vertices
    \node[draw,circle,cross,minimum width=8pt] at (AB) {};
    \fill (CD) circle (4pt);
    % Add labels
    \node[below=4pt] at (AB) {$AB$};
    \node[below=4pt] at (CD) {$CD$};
  \end{tikzpicture} \sim \frac{1}{\twbr{CD12}\twbr{CD23}\twbr{CD34}\twbr{CD41}\twbr{ABCD}}\,,
  \label{eq:oneloop}
\end{equation}
where we have excluded the four propagators $\twbr{AB\,i\,i{+}1}$ that do not participate. We only care about Landau diagrams which include $\twbr{ABCD}$, since this is the only one that involves $AB$ and hence can potentially impose a non-trivial condition on $z$. Let's work our way down from the most constraining topologies. Starting with the pentagon, which includes all possible propagators:

\begin{equation}
    \twbr{CD12}=\twbr{CD23}=\twbr{CD34}=\twbr{CD14}=\twbr{ABCD}=0\,,
\end{equation}
we see that the solution is overconstrained: the first four conditions already fix the loop line to either $CD=(13)$ or $CD=(24)$, which then implies that one of $\twbr{AB13}$ or $\twbr{AB24}$ must vanish. However, as emphasized in Section~\ref{sec: landau} the cut conditions should be satisfied for generic external kinematics, so we can neglect this configuration.

Next, we look at box topologies. Without loss of generality, we can choose e.g. the propagators
\begin{equation}
\twbr{CD12}=\twbr{CD23}=\twbr{CD34}=\twbr{ABCD}=0\,,
\end{equation}
which corresponds to the diagram
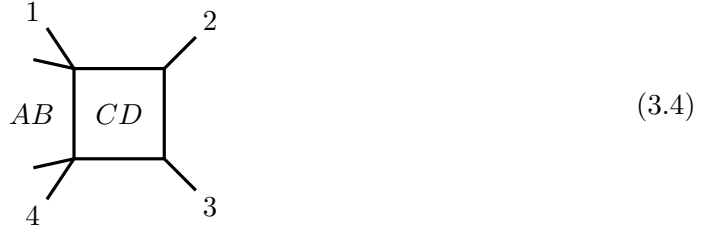
\begin{equation}
  \begin{tikzpicture}[baseline=(current bounding box.center), thick, scale=0.6]
    % Define coordinates for the corners of the box
    \coordinate (BL) at (0,0);
    \coordinate (TL) at (0,2);
    \coordinate (TR) at (2,2);
    \coordinate (BR) at (2,0);

    % Draw the main box
    \draw[very thick] (BL) -- (TL) -- (TR) -- (BR) -- cycle;

    % Top-left external lines (shortened)
    \draw[very thick] (TL) -- +(-0.6, 0.9) node[above left, inner sep=2pt] {$1$};
    \draw[very thick] (TL) -- +(-0.9, 0.2);

    % Top-right external line (shortened)
    \draw[very thick] (TR) -- +(0.7, 0.7) node[above right, inner sep=2pt] {$2$};

    % Bottom-right external line (shortened)
    \draw[very thick] (BR) -- +(0.7, -0.7) node[below right, inner sep=2pt] {$3$};

    % Bottom-left external lines (shortened)
    \draw[very thick] (BL) -- +(-0.6, -0.9) node[below left, inner sep=2pt] {$4$};
    \draw[very thick] (BL) -- +(-0.9, -0.2);

    % Labels
    \node at (1, 1) {$CD$};
    \node[left] at (-0.2, 1) {$AB$};
  \end{tikzpicture}
\end{equation}
This is a two-mass-hard Schubert problem, which can be readily solved (see e.g.~\cite{Arkani-Hamed:2010pyv}). The two branches are:
\begin{equation}
\begin{aligned}
        CD &= (123)\cap(3AB) ~{\rm or}\\
        CD &= (234)\cap(2AB)\,.
\end{aligned}
\end{equation}
Meanwhile, the pinch condition
\begin{equation}
    \alpha_2(y_{CD}-x_2)^\mu + \alpha_3(y_{CD}-x_3)^\mu + \alpha_4(y_{CD}-x_4)^\mu + \alpha_0(y_{CD} - y_{AB})^\mu =0
\end{equation}
can be projected onto each of the propagators, which yields the following matrix equation:
\begin{equation}
    \begin{pmatrix}
        0&0&\twbr{1234}&\twbr{AB12} \\
        0&0&0&\twbr{AB23}\\
        \twbr{1234}&0&0&\twbr{AB34}\\
        \twbr{AB12}&\twbr{AB23}&\twbr{AB34}&0
    \end{pmatrix}\begin{pmatrix}
        \alpha_2\\\alpha_3\\\alpha_4\\\alpha_0
    \end{pmatrix}=0\,.
\end{equation}
Here, we have used relations such as
\begin{align}
    (y_{CD}{-}x_2)\cdot(y_{CD}{-}y_{AB}) &= -\frac{1}{2}\left[ (y_{CD}{-}x_2{-}y_{CD}{+}y_{AB})^2{-}(y_{CD}{-}x_2)^2 - (y_{CD}{-}y_{AB})^2 \right]\nonumber\\ &= -\frac{1}{2}(y_{AB}{-}x_2)^2\,,
\end{align}
and the fact that $(x_i - x_j)^2 \propto \twbr{i\,i{+}1\,j\,j{+}1}$, up to some brackets involving the infinity bitwistor, which cancel out in any determinant-like condition.

Solutions with all Schwinger parameters being non-zero exist only if $\langle 1234 \rangle = \langle AB 23\rangle =0$. Since we only want to impose constraints on the frozen loop line $AB$, we do not consider this a valid singularity for our purposes. Therefore we move on to the next case, which is a triangle diagram, for which there are two inequivalent choices of propagators. First we consider
\begin{align}
    \twbr{CD12}=\twbr{CD23}=\twbr{ABCD}=0
\end{align}
which corresponds to the Landau diagram
\begin{gather}
    \begin{tikzpicture}[baseline=(current bounding box.center), thick, scale=0.7]
    % Define coordinates for the corners of the triangle
    \coordinate (BL) at (0,0);
    \coordinate (BR) at (2,0);
    \coordinate (T)  at (1, 1.732); % Height of an equilateral triangle with base 2
    % Draw the main triangle
    \draw[very thick] (BL) -- (T) -- (BR) -- cycle;
    % Top external line
    \draw[very thick] (T) -- +(0, 0.9) node[above, inner sep=2pt] {$2$};
    % Bottom-left external lines
    \draw[very thick] (BL) -- +(-0.9, -0.2)node[left, inner sep=2pt] {$1$};
    \draw[very thick] (BL) -- +(-0.6, -0.8) ;
    % Bottom-right external lines
    \draw[very thick] (BR) -- +(0.9, -0.2) node[right, inner sep=2pt] {$3$};
    \draw[very thick] (BR) -- +(0.6, -0.8) ;
    % Labels
    \node at (1, 0.65) {$CD$};
    \node[below] at (1, -0.1) { $AB$};
    \end{tikzpicture}
\end{gather}
Meanwhile, the pinch condition is written in matrix form as:
\begin{equation}
    \begin{pmatrix}
        0&0&\twbr{AB12} \\
        0&0&\twbr{AB23}\\
        \twbr{AB12}&\twbr{AB23}&0
    \end{pmatrix}\vec{\alpha}=0\,
\end{equation}
Demanding that all three Schwinger parameters are non-zero imposes $\twbr{AB12}=\twbr{AB23}=0$, which is a codimension-2 locus in kinematic space and hence can't correspond to a branch point in $z$.

A second type of triangle configuration is given by the  cut conditions
\begin{gather}
    \twbr{CD12}=\twbr{CD34}=\twbr{ABCD}=0\\
    \begin{tikzpicture}[baseline=(current bounding box.center), thick, scale=0.7]
    % Define coordinates for the corners of the triangle
    \coordinate (BL) at (0,0);
    \coordinate (BR) at (2,0);
    \coordinate (T)  at (1, 1.732); % Height of an equilateral triangle with base 2
    % Draw the main triangle
    \draw[very thick] (BL) -- (T) -- (BR) -- cycle;
    % Top external lines
    \draw[very thick] (T) -- +(-0.4, 0.7) node[above left, inner sep=2pt] {$2$};
    \draw[very thick] (T) -- +(0.4, 0.7) node[above right, inner sep=2pt] {$3$};
    % Bottom-left external lines
    \draw[very thick] (BL) -- +(-0.9, -0.2) node[left, inner sep=2pt] {$1$};
    \draw[very thick] (BL) -- +(-0.6, -0.8);
    % Bottom-right external lines
    \draw[very thick] (BR) -- +(0.9, -0.2) node[right, inner sep=2pt] {$4$};
    \draw[very thick] (BR) -- +(0.6, -0.8);
    % Labels
    \node at (1, 0.65) {$CD$};
    \node[below] at (1, -0.1) {$AB$};
    \end{tikzpicture}
\end{gather}
and the corresponding pinch condition has the form
\begin{equation}
    \begin{pmatrix}
        0&\twbr{1234}&\twbr{AB12} \\
        \twbr{1234}&0&\twbr{AB34}\\
        \twbr{AB12}&\twbr{AB34}&0
    \end{pmatrix}\vec{\alpha}=0\,,
\end{equation}
which also doesn't have a nontrivial solution at codimension-1.

Finally we have the bubble topology, involving only two cut conditions, for example:
\begin{gather}
\twbr{CD12}=\twbr{ABCD}=0\\
    \begin{tikzpicture}[baseline=(current bounding box.center), thick, scale=0.7]
    % Define the two main vertices
    \coordinate (L) at (-1.8, 0);
    \coordinate (R) at (1.8, 0);
    % Draw the four internal lines (arcs)
    \draw[very thick] (L) to[bend left=45] (R);
    \draw[very thick] (L) to[bend right=45] (R);
    % Left external lines
    \draw[very thick] (L) -- +(-0.8, 0.8);
    \draw[very thick] (L) -- +(-0.8, -0.8) node[below left, inner sep=2pt] {$2$};
    % Right external lines
    \draw[very thick] (R) -- +(0.8, 0.8);
    \draw[very thick] (R) -- +(0.8, -0.8) node[below right, inner sep=2pt] {$1$};
    % Labels
    \node at (0, 1.1) {$AB$};
    \node at (0, 0) {$CD$};
    \end{tikzpicture}
\end{gather}
The pinch condition now involves a $2\times2$ equation of the form
\begin{equation}
    \begin{pmatrix}
        0&\twbr{AB12} \\
        \twbr{AB12}&0
    \end{pmatrix}\vec{\alpha}=0
\end{equation}
and clearly has a non-trivial solution only if $\twbr{AB12}=0$. This corresponds to a singularity $z=0$, and consideration of the cyclically permuted diagram would reveal a singularity at $\twbr{AB23} = 0$ which is at $z = \infty$.

In Section~\ref{sec: landau} we alerted the reader to a subtlety in our algorithm regarding how to properly account for branches of singularities of the Landau equations for which both $q_e^2 = 0$ and $\alpha_e = 0$ for some edge(s) $e$. At one-loop this only occurs for three-mass or two-mass-easy box topologies, which we have not encountered as they do not appear in~(\ref{eq:oneloop}).

This exhausts all the possible Landau diagrams at one loop, and thus we conclude that the only singularities are at $z=0,\infty$. This is in accordance with the actual integrated result for the negative geometry~\cite{Alday:2012hy} which is given by the simple function
\begin{equation}
    F^{(1)} = \log^2z+\pi^2\,.
\end{equation}
Moreover, since any other negative geometry associated with a tree graph can be related to this one by applying some sequence of Laplacian differential operators~\cite{Arkani-Hamed:2021iya}, it follows that $z=0,\infty$ are the only singularities of any such negative geometries.

\subsection{Two-loop negative geometry}\label{sec: two loop analysis}

We now move on to the two-loop example, which is the first case that corresponds to a one-cycle topology. Due to the large number of possible cut conditions, we will only look at a couple of specific Landau diagrams to illustrate our methods.  The multiloop propagators present in the integrand are
\begin{equation}
    \begin{tikzpicture}[baseline={([yshift=-1.5ex]current bounding box.center)}, 
    thick, scale=0.9,
    cross/.style={path picture={
      \draw[black] (path picture bounding box.south east) -- (path picture bounding box.north west) (path picture bounding box.south west) -- (path picture bounding box.north east);
    }}]
    \coordinate (AB) at (0,0);
    \coordinate (AB1) at (4.25pt,4.25pt);
    \coordinate (AB2) at (6pt,0);
    \coordinate (CD) at (0.875,1.25);
    \coordinate (EF) at (1.75,0);
    % Draw the square
    \draw[very thick, Maroon] (AB1) -- (CD) -- (EF) -- (AB2);
    % Draw the vertices
    \node[draw,circle,cross,minimum width=8pt] at (AB) {};
    \fill (CD) circle (4pt);
    \fill (EF) circle (4pt);
    % Add labels
    \node[below=4pt] at (AB) {$AB$};
    \node[above=4pt] at (CD) {$CD$};
    \node[below=4pt] at (EF) {$EF$};
  \end{tikzpicture} \sim \frac{1}{\twbr{ABCD}\twbr{CDEF}\twbr{ABEF}}\,.
\end{equation}
The first example we will look at is the double pentagon with the following cut conditions:
\begin{gather}
    \begin{aligned}
        \twbr{CD12}&=\twbr{CD14}=\twbr{CD34}=\twbr{ABCD}=0\,,\\
        \twbr{EF12}&=\twbr{EF23}= \twbr{EF34}= \twbr{ABEF} =\twbr{CDEF}=0\,,
    \end{aligned}
    \end{gather}
    corresponding to the Landau diagram
    \begin{gather}
    \begin{tikzpicture}[baseline=(current bounding box.center), scale=0.4]
    % --- Shared Vertices ---
    \coordinate (T) at (0, 1.2);
    \coordinate (B) at (0, -1.2);
    % --- Left Pentagon Vertices ---
    \coordinate (TL) at (-2.2, 2.2);
    \coordinate (ML) at (-3.8, 0);
    \coordinate (BL) at (-2.2, -2.2);
    % --- Right Pentagon Vertices ---
    \coordinate (TR) at (2.2, 2.2);
    \coordinate (MR) at (3.8, 0);
    \coordinate (BR) at (2.2, -2.2);
    % --- Internal Edges ---
    \draw[very thick] (T) -- (B);
    \draw[very thick] (T) -- (TL) -- (ML) -- (BL) -- (B);
    \draw[very thick] (T) -- (TR) -- (MR) -- (BR) -- (B);
    % --- External Legs - Left ---
    \draw[very thick] (TL) -- +(-0.6, 1.0) node[above left, inner sep=1pt] { $1$};
    \draw[very thick] (ML) -- +(-1.0, 0) node[left, inner sep=2pt] { $4$};
    % Bottom Left V-legs (massive corner)
    \draw[very thick] (BL) -- +(-0.6, -1.0) node[below left, inner sep=1pt] { $3$};
    \draw[very thick] (BL) -- +(0, -1.1);
    % --- External Legs - Right ---
    \draw[very thick] (TR) -- +(0.6, 1.0) node[above right, inner sep=1pt] { $2$};
    \draw[very thick] (MR) -- +(1.0, 0) node[right, inner sep=2pt] { $3$};
    % Bottom Right V-legs (massive corner)
    \draw[very thick] (BR) -- +(0, -1.1);
    \draw[very thick] (BR) -- +(0.6, -1.0) node[below right, inner sep=1pt] { $4$};
    % --- Internal Labels ---
    \node at (-1.8, 0) { $CD$};
    \node at (1.8, 0) { $EF$};
    \node at (0, -2.2) { $AB$};
  \end{tikzpicture}\,
\end{gather}
The pinch condition for the line $CD$ can be written as the matrix equation
\begin{equation}
    \begin{pmatrix}
        0&0&\twbr{1234}&\twbr{AB12}&0\\
        0&0&0&\twbr{AB14}&\twbr{EF14}\\
        \twbr{1234}&0&0&\twbr{AB34}&0\\
        \twbr{AB12}&\twbr{AB14}&\twbr{AB34}&0&0\\
        0&\twbr{EF14}&0&0&0
    \end{pmatrix}
    \vec{\alpha} =0\,,
\end{equation}
which has no solution if we demand that all Schwinger proper times are non-zero. In fact, one can show that this is the case for any permutation of the double pentagon topology, and thus we don't see any (leading) Landau singularities from these diagrams.

The next example of a diagram we consider is a double box with the following cut conditions:
\begin{gather}
\begin{aligned}
    &\twbr{CD34}=\twbr{CD14}=\twbr{ABCD}=0\,,\\
    &\twbr{EF12}=\twbr{EF23}=\twbr{ABEF}=\twbr{CDEF}=0\,,
\end{aligned}\\
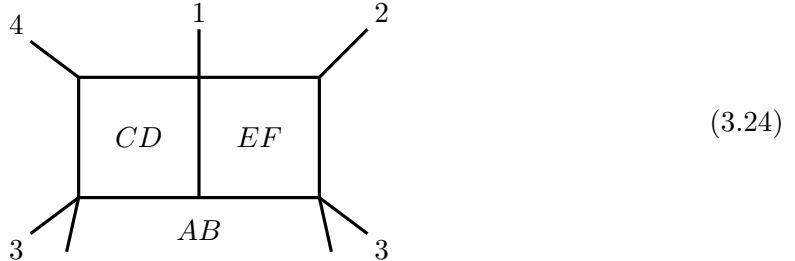
    \begin{tikzpicture}[baseline=(current bounding box.center), thick, scale=0.8]
    % Define coordinates for the corners and midpoints
    \coordinate (BL) at (0,0);
    \coordinate (Bmid) at (2,0);
    \coordinate (BR) at (4,0);
    \coordinate (TL) at (0,2);
    \coordinate (Tmid) at (2,2);
    \coordinate (TR) at (4,2);
    % Draw the double box (outer frame and inner divider)
    \draw[very thick] (BL) -- (BR) -- (TR) -- (TL) -- cycle;
    \draw[very thick] (Bmid) -- (Tmid);
    % Top-left external line
    \draw[very thick] (TL) -- +(-0.8, 0.6) node[above left, inner sep=2pt] {$4$};
    % Top-middle external line
    \draw[very thick] (Tmid) -- +(0, 0.8) node[above, inner sep=2pt] {$1$};
    % Top-right external line
    \draw[very thick] (TR) -- +(0.8, 0.8) node[above right, inner sep=2pt] {$2$};
    % Bottom-right external lines (two legs)
    \draw[very thick] (BR) -- +(0.8, -0.6) node[below right, inner sep=2pt] {$3$};
    \draw[very thick] (BR) -- +(0.2, -0.9);
    % Bottom-left external lines (two legs)
    \draw[very thick] (BL) -- +(-0.8, -0.6) node[below left, inner sep=2pt] {$3$};
    \draw[very thick] (BL) -- +(-0.2, -0.9);
    % Internal Labels
    \node at (1, 1) {$CD$};
    \node at (3, 1) {$EF$};
    % External Label
    \node[below] at (2, -0.2) {$AB$};
    \end{tikzpicture}
\end{gather}
We work in a loop-by-loop approach, first focusing on the line $EF$, which means that we start by treating both $AB$ and $CD$ as if they were completely arbitrary, external kinematics. Thus, for the purpose of solving $EF$ we are effectively looking at the three-mass box problem
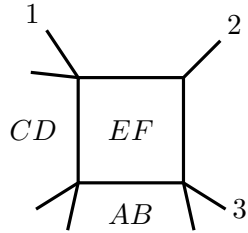
\begin{equation*}
  \begin{tikzpicture}[baseline=(current bounding box.center), thick, scale=0.7]
    % Define coordinates for the corners of the box
    \coordinate (BL) at (0,0);
    \coordinate (TL) at (0,2);
    \coordinate (TR) at (2,2);
    \coordinate (BR) at (2,0);

    % Draw the main box
    \draw[very thick] (BL) -- (TL) -- (TR) -- (BR) -- cycle;

    % Top-left external lines
    \draw[very thick] (TL) -- +(-0.6, 0.9) node[above left, inner sep=2pt] {$1$};
    \draw[very thick] (TL) -- +(-0.9, 0.1);

    % Top-right external line
    \draw[very thick] (TR) -- +(0.7, 0.7) node[above right, inner sep=2pt] {$2$};

    % Bottom-right external lines
    \draw[very thick] (BR) -- +(0.8, -0.5) node[right, inner sep=2pt] {$3$};
    \draw[very thick] (BR) -- +(0.2, -0.9);

    % Bottom-left external lines
    \draw[very thick] (BL) -- +(-0.8, -0.5);
    \draw[very thick] (BL) -- +(-0.2, -0.9);

    % Labels
    \node at (1, 1) {$EF$};
    \node[left] at (-0.2, 1) {$CD$};
    \node[below] at (1, -0.2) {$AB$};
  \end{tikzpicture}
\end{equation*}
which has two solutions \cite{Arkani-Hamed:2010pyv}
\begin{equation}
\begin{aligned}
    EF &= \left( (123)\cap AB,\, CD\cap(123) \right) ~ {\rm or}\\
    EF &= (2AB)\cap(CD2)=(2,AB\cap CD)\,,
\end{aligned}
\end{equation}
where the last expression has been simplified using the fact that $AB$ and $CD$ will be imposed to intersect \textit{a posteriori}. Note that the first solution corresponds to the branch where $EF$ lies in the $(123)$ plane, while the second solution corresponds to $EF$ passing through $Z_2$.

Next, we derive the pinch condition for $EF$. In terms of four-vectors:
\begin{equation}
    \alpha_2(y_{EF} - x_2)^\mu + \alpha_3(y_{EF} - x_3)^\mu + \alpha_0(y_{EF}-y_{AB})^\mu + \beta(y_{EF}-y_{CD})^\mu = 0\,.
\end{equation}
Projecting this onto the four propagators, we obtain a matrix equation of the form
\begin{equation}
    \begin{pmatrix}
        0&0&\twbr{AB12}&\twbr{CD12}\\
        0&0&\twbr{AB23}&\twbr{CD23}\\
        \twbr{AB12}&\twbr{AB23}&0&0\\
        \twbr{CD12}&\twbr{CD23}&0&0
    \end{pmatrix}
    \vec{\alpha} =0\,.
\end{equation}
A non-trivial solution exists only if the determinant vanishes, which imposes
\begin{equation}
    \twbr{CD12}\twbr{AB23}-\twbr{CD23}\twbr{AB12}=0\,.
\end{equation}
Using \eqref{eq:plane int id}, this can be recast as
\begin{equation}
    \twbr{1\,2\,3\,AB\cap CD}=0\,.
\end{equation}
In other words, the intersection point of the lines $AB$ and $CD$ has to lie on the $(123)$ plane.
Since this is a constraint on the line $CD$, we can use it in conjunction with the remaining three cut conditions to write four conditions on the loop line $CD$:
\begin{equation}
    \twbr{CD34}=\twbr{CD14}=\twbr{ABCD} = \twbr{1\,2\,3\,AB\cap CD}=0\,.
\end{equation}
Again, there are two solutions to this Schubert problem. The first one has $CD$ lying on the (341) plane. However, since $(123)\cap(341)=(13)$, from the last condition we must have that $AB\cap CD\in (1,3)$. Thus we have $\twbr{AB13}=0$, which establishes a singularity at $z=-1$.

The second possibility is that $CD$ passes through the point $Z_4$. Applying the last two cut conditions fully fixes
\begin{equation}
    CD = (\twbr{AB23}1 - \twbr{AB13}2 + \twbr{AB12}3,\, 4) = (AB\cap(123),\,4)\,.
\end{equation}
Lastly, the pinch condition for $CD$ has the form
\begin{equation}
    \begin{pmatrix}
        0&0&\twbr{AB34}&\twbr{EF34}\\
        0&0&\twbr{AB14}&\twbr{EF14}\\
        \twbr{AB34}&\twbr{AB14}&0&0\\
        \twbr{EF34}&\twbr{EF14}&0&0
    \end{pmatrix}
    \vec{\alpha} =0\,.
\end{equation}
Again we need the $2\times 2$ determinant to vanish:
\begin{equation}
      \twbr{EF34}\twbr{AB14}-\twbr{EF14}\twbr{AB34}=\twbr{3\,4\,1\,AB\cap EF}=0\,,
\end{equation}
i.e. the intersection point of $AB$ and $EF$ must lie on the $(341)$ plane.

This pinch condition now has to be inserted back into our solutions for $EF$ so that we obtain a constraint on the external kinematics. Now, if we consider the first solution we derived,
\begin{equation}
    EF = \left( (123)\cap AB,\, CD\cap(123) \right),
\end{equation}
we see that it is not really well defined; from the first pinch condition it is clear that both points are the same and don't actually define a line. Thus, we need to consider the second solution
\begin{equation}
    EF = (2AB)\cap(CD2) = (2,AB\cap CD)\,.
\end{equation}
Now, we know that $AB\cap CD\in (123)$, and the other pinch condition tells us that it also has to lie on the $(341)$ plane. Thus, we must have $\twbr{AB13}=0$, so we conclude there is a singularity at $z=-1$.

As a last comment, note that the final solution for the loop lines has $CD$ both lying on $(341)$ \emph{and} passing through $Z_4$, and similarly for the line $EF$. In theory, this would correspond to a collinear configuration, which is not allowed by the negative geometry. However, in this case we cannot discard the solution, since it is only localized \emph{after} imposing conditions on $AB$, meaning the monodromy around the pole $z=-1$ is non-trivial, and thus needs to be considered as an actual Landau singularity.

To illustrate the fact that there is nothing else at 2 loops, consider another example of a double-box, this time with a four-mass problem with cut conditions
\begin{gather}
    \twbr{CD12}=\twbr{CD34}=\twbr{EF23}=\twbr{EF14}= \twbr{ABCD}=\twbr{ABEF}=\twbr{CDEF}=0\,,\\
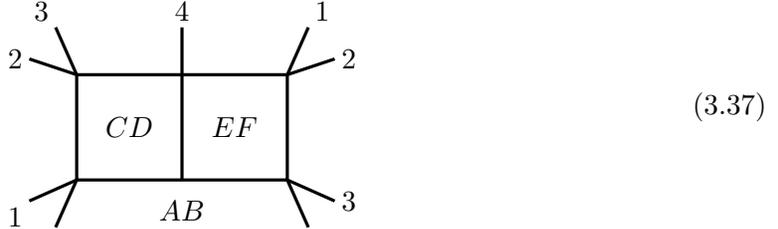
    \begin{tikzpicture}[baseline=(current bounding box.center), thick, scale=0.7]
    % Define coordinates for the corners and midpoints
    \coordinate (BL) at (0,0);
    \coordinate (Bmid) at (2,0);
    \coordinate (BR) at (4,0);
    \coordinate (TL) at (0,2);
    \coordinate (Tmid) at (2,2);
    \coordinate (TR) at (4,2);
    % Draw the double box (outer frame and inner divider)
    \draw[very thick] (BL) -- (BR) -- (TR) -- (TL) -- cycle;
    \draw[very thick] (Bmid) -- (Tmid);
    % Top-left external lines
    \draw[very thick] (TL) -- +(-0.9, 0.3) node[left, inner sep=2pt] {$2$};
    \draw[very thick] (TL) -- +(-0.4, 0.9) node[above left, inner sep=2pt] {$3$};
    % Top-middle external line
    \draw[very thick] (Tmid) -- +(0, 0.9) node[above, inner sep=2pt] {$4$};
    % Top-right external lines
    \draw[very thick] (TR) -- +(0.4, 0.9) node[above right, inner sep=2pt] {$1$};
    \draw[very thick] (TR) -- +(0.9, 0.3) node[right, inner sep=2pt] {$2$};
    % Bottom-right external lines
    \draw[very thick] (BR) -- +(0.9, -0.4) node[right, inner sep=2pt] {$3$};
    \draw[very thick] (BR) -- +(0.4, -0.9);
    % Bottom-left external lines
    \draw[very thick] (BL) -- +(-0.9, -0.4)node[below left, inner sep=2pt] {$1$};
    \draw[very thick] (BL) -- +(-0.4, -0.9) ;
    % Internal Labels
    \node at (1, 1) {$CD$};
    \node at (3, 1) {$EF$};
    % External Label
    \node[below] at (2, -0.2) {$AB$};
    \end{tikzpicture}
\end{gather}
For this case, we don't even need to look at the explicit solutions for the loop lines (which are quite complicated). Indeed, already for the pinch condition of e.g. $CD$,
\begin{equation}
    \begin{pmatrix}
        0&\twbr{1234}&\twbr{AB12}&\twbr{EF12}\\
        \twbr{1234}&0&\twbr{AB34}&\twbr{EF34}\\
        \twbr{AB12}&\twbr{AB34}&0&0\\
        \twbr{EF12}&\twbr{EF34}&0&0
    \end{pmatrix}
    \vec{\alpha} = 0\,,
\end{equation}
we see that there is no possible solution where all Schwinger parameters are non-zero if we demand that the external kinematics remain generic (specifically, if $\twbr{1234}\neq0$).

Having studied some examples in detail, we now proceed with a general recursive argument that works for Landau diagrams at arbitrary loop order, and reduces the need to look at many individual cases.

\section{Recursive proof at arbitrary loop order}\label{sec: general claims}

In this section we show that Landau diagrams associated with the one-cycle negative geometry~(\ref{eq:onecycle}) only give rise to singularities at $z\in \{-1,0,\infty\}$. The argument is recursive and relies mainly on the constraints imposed on the first few loop lines in the cycle. First, we argue in Section~\ref{sec: bubbles and triangles} that all bubble and triangle sub-diagrams can be dropped. Then in Section~\ref{sec: hexagon or pentagon} we consider the case when the first loop line $AB_1$ is a hexagon or pentagon. Finally, we discuss the cases where the first loop line is a four-mass box in Section \ref{sec: 4mass box} and a three-mass box in Section~\ref{sec: 3mass box}. Recursive approaches to Landau analysis, similar to the one we present here, have been used in other work including~\cite{Correia:2021etg,Caron-Huot:2024brh,Hollering:2026gjz,Hollering:2026rlv}.

An important simplification in our analysis is that we only need to study Landau diagrams that include edges corresponding to \emph{all} consecutive pairs of loop lines $\langle AB_\ell AB_{\ell+1} \rangle$, since any other diagram will have singularities associated with products of tree-graph negative geometries, which only have singularities at $z \in \{0,\infty\}$ as we have seen in Section~\ref{sec: one-loop analysis}.

\subsection{Elimination of bubbles and triangles}\label{sec: bubbles and triangles}

Firstly, we show that any Landau diagram containing a bubble or a triangle as a subgraph is completely equivalent to a lower-loop problem where that loop line is simply removed.
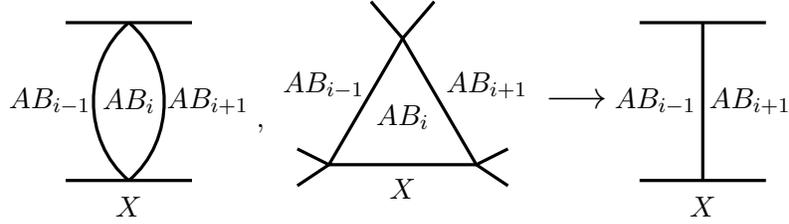
\begin{figure}
\begin{center}
  \begin{tikzpicture}[baseline=(current bounding box.center), thick, scale=0.7]
    
    % ==========================================
    % Left Diagram (Bubble)
    % ==========================================
    % Top and bottom horizontal lines
    \draw[very thick] (-1.2, 1.5) -- (1.2, 1.5);
    \draw[very thick] (-1.2, -1.5) -- (1.2, -1.5);
    
    % The bubble arcs connecting top and bottom
    \draw[very thick] (0, -1.5) to[bend left=50] (0, 1.5);
    \draw[very thick] (0, -1.5) to[bend right=50] (0, 1.5);

    % Labels
    \node at (-1.5, 0) {$AB_{i-1}$};
    \node at (0, 0) {$AB_i$};
    \node at (0, -2) {$X$};
    \node at (1.5, 0) {$AB_{i+1}$};

    % Comma separator
    \node at (2.5, -0.5) {,};

    % ==========================================
    % Middle Diagram (Triangle)
    % ==========================================
    % Triangle vertices
    \coordinate (T) at (5.2, 1.2);
    \coordinate (BL) at (3.8, -1.2);
    \coordinate (BR) at (6.6, -1.2);

    % Triangle edges
    \draw[very thick] (T) -- (BL) -- (BR) -- cycle;

    % External legs
    \draw[very thick] (T) -- (4.6, 1.9);
    \draw[very thick] (T) -- (5.8, 1.9);
    \draw[very thick] (BL) -- (3.2, -0.9);
    \draw[very thick] (BL) -- (3.2, -1.6);
    \draw[very thick] (BR) -- (7.2, -0.9);
    \draw[very thick] (BR) -- (7.2, -1.6);

    % Labels
    \node at (3.7, 0.3) {$AB_{i-1}$};
    \node at (5.2, -0.3) {$AB_i$};
    \node at (6.8, 0.3) {$AB_{i+1}$};
    \node at (5.2, -1.65) {$X$};

    % ==========================================
    % Tilde Operator
    % ==========================================
    \node at (8.5, 0) {\Large ${\longrightarrow}$};

    % ==========================================
    % Right Diagram (Vertical Line)
    % ==========================================
    % Top and bottom horizontal lines
    \draw[very thick] (9.7, 1.5) -- (12.1, 1.5);
    \draw[very thick] (9.7, -1.5) -- (12.1, -1.5);
    
    % The straight vertical line connecting top and bottom
    \draw[very thick] (10.9, -1.5) -- (10.9, 1.5);

    % Labels
    \node at (10, 0) {$AB_{i-1}$};
    \node at (11.8, 0) { $AB_{i+1}$};
    \node at (10.9, -2) {$X$};

  \end{tikzpicture}
  \end{center}
  \caption{Landau diagrams with a bubble or triangle inside them are equivalent to a subgraph where that specific loop line has been removed.}
\label{fig: bubble triangle}
\end{figure}
The argument is very simple \cite{Dennen:2016mdk}. Let's start with the bubble case; the cut and pinch conditions read
\begin{equation}
    \begin{aligned}
        (y_{i-1} - y_i)^2 = (y_{i+1}-y_i)^2 &=  0\,,\\
        \beta_{i-1}(y_{i-1} - y_i)^\mu + \beta_{i+1}(y_{i+1} - y_i)^\mu &= 0\,.    \end{aligned}
\end{equation}
Projecting the pinch condition onto the propagators, the result is $\twbr{AB_{i-1}AB_{i+1}}=0$, i.e.~the lines $AB_{i-1}$ and $AB_{i+1}$ cut each other, which would be the cut condition of the problem where the region corresponding to $AB_i$ is removed from the graph. Moreover, explicitly solving for $y_i^\mu$, we get:
\begin{equation}
    \begin{aligned}
        y_i^\mu = \frac{1}{\beta_{i-1}+\beta_{i+1}}\left( \beta_{i-1}\,y_{i-1}^\mu + \beta_{i+1}\,y_{i+1}^\mu \right).
    \end{aligned}
\end{equation}
Thus, the propagators themselves read
\begin{equation}
    \begin{aligned}
        \beta_{i-1}(y_{i-1} - y_i)^\mu = \frac{\beta_{i-1}\beta_{i+1}}{\beta_{i-1}+\beta_{i+1}}(y_{i-1} - y_{i+1})^\mu = \beta'(y_{i-1} - y_{i+1})^\mu\,,\\
        \beta_{i+1}(y_{i+1} - y_i)^\mu = \frac{\beta_{i-1}\beta_{i+1}}{\beta_{i-1}+\beta_{i+1}}(y_{i+1} - y_{i-1})^\mu = \beta'(y_{i+1} - y_{i-1})^\mu\,.
    \end{aligned}
\end{equation}
As a result, the pinch condition for e.g.~the loop line $AB_{i-1}$ can be written as
\begin{equation}
    \sum_{\text{other conditions}} (\cdots)\ + \beta_{i-1}(y_{i-1}-y_i)^\mu = \sum_{\text{other conditions}} (\cdots)\ + \beta'(y_{i-1}-y_{i+1})^\mu=0\,,
\end{equation}
and similarly for $AB_{i+1}$. However, those are simply the pinch conditions that we would obtain if the loop line $AB_i$ were absent (see Figure~\ref{fig: bubble triangle}), so we have proven that the problem in its entirety just reduces to the lower-loop case.

A very similar argument holds for triangles. Now we have
\begin{equation}
    \begin{aligned}
        (x - y_i)^2=(y_{i-1} - y_i)^2 = (y_{i+1}-y_i)^2 &=  0\,,\\
        \alpha_x(x - y_i)^\mu+\beta_{i-1}(y_{i-1} - y_i)^\mu + \beta_{i+1}(y_{i+1} - y_i)^\mu &= 0\,,
    \end{aligned}
\end{equation}
where $x^\mu \sim X^{IJ}$ is some arbitrary external line (corresponding to the bottom line of the triangle in Figure~\ref{fig: bubble triangle}). Solving the pinch condition for $AB_i$ gives the constraints
\begin{equation}
    \twbr{AB_{i-1}X} = \twbr{AB_{i+1}X} = \twbr{AB_{i-1}AB_{i+1}}=0\,,
\end{equation}
which are the same cut conditions as if $AB_i$ were removed from the diagram (as evident in the figure). Again, solving explicitly for the $AB_i$ loop line yields
\begin{equation}
    y_i^\mu = \frac{1}{\alpha_x + \beta_{i-1}+\beta_{i+1}}\left(\alpha_x x^\mu+ \beta_{i-1}\,y_{i-1}^\mu + \beta_{i+1}\,y_{i+1}^\mu \right),
\end{equation}
so that
\begin{equation}
    \begin{aligned}
        \beta_{i-1}(y_{i-1} - y_i)^\mu &= \frac{\beta_{i-1}}{\alpha_x+\beta_{i-1}+\beta_{i+1}}\left[\alpha_x(y_{i-1}-x)^\mu + \beta_{i+1}(y_{i-1} - y_{i+1})^\mu\right] \\&= \alpha_x^{(i-1)}(y_{i-1}-x)^\mu+\beta'(y_{i-1} - y_{i+1})^\mu\,,\\
        \beta_{i+1}(y_{i+1} - y_i)^\mu &= \frac{\beta_{i+1}}{\alpha_x+\beta_{i-1}+\beta_{i+1}}\left[\alpha_x(y_{i+1}-x)^\mu + \beta_{i-1}(y_{i+1} - y_{i-1})^\mu\right] \\ &= \alpha_x^{(i+1)}(y_{i+1}-x)^\mu+ \beta'(y_{i+1} - y_{i-1})^\mu\,.
    \end{aligned}
\end{equation}
Inserting this into the pinch conditions for $AB_{i\pm1}$, this will become (upon an overall rescaling of the Schwinger parameters) the pinch condition for the problem where $AB_i$ is removed and the cuts $\twbr{AB_{i\pm1}X}=0$ are potentially added (depending on whether they were there before or not).

All in all, we conclude that the problem associated with any Landau diagram involving a bubble or triangle subgraph can be reduced to the one where that specific loop line has been removed, and its cuts with the external and internal kinematics have been transferred to adjacent loops. As a result, we only need to consider those cases where all regions are boxes, pentagons or hexagons.

\subsection{Hexagon or pentagon as the first loop line}\label{sec: hexagon or pentagon}

We will now show that all cases where the first loop line $AB_1$ is associated with a hexagon or a pentagon lead to singularities corresponding to $z \in \{-1,0,\infty\}$.
\begin{figure}
\centering
  \begin{tikzpicture}[baseline=(current bounding box.center), scale=0.9]
    
    % Define a smaller shared side length for compactness
    \def\a{1.1}
    \def\b{1.6}
    \def\Rp{1.1} % Radius for regular pentagon: a / (2 * sin(36))
    
    % Spread out the centers to ensure absolutely no overlap
    \def\Xpent{6} % Center X of pentagon
    \def\Xbox{11.5} % Center X of box

    % ==========================================
    % DIAGRAM 1: Regular Hexagon
    % ==========================================
    \coordinate (H1) at (90:\a);
    \coordinate (H2) at (150:\a);
    \coordinate (H3) at (210:\a);
    \coordinate (H4) at (270:\a);
    \coordinate (H5) at (330:\a);
    \coordinate (H6) at (30:\a);
    
    \draw[very thick] (H1) -- (H2) -- (H3) -- (H4) -- (H5) -- (H6) -- cycle;
    
    % External Legs
    % Top-left V (1)
    \draw[very thick] (H2) -- +(-0.65, 0.7) node[above left, inner sep=1pt] {\large $1$};
    \draw[very thick] (H2) -- +(-0.85, 0.2);
    % Bottom-left V (1)
    \draw[very thick] (H3) -- +(-0.65, -0.7) node[below left, inner sep=1pt] {\large $1$};
    \draw[very thick] (H3) -- +(-0.85, -0.2);
    % Top (2)
    \draw[very thick] (H1) -- +(0, 0.8) node[above, inner sep=1pt] {\large $2$};
    % Bottom (4)
    \draw[very thick] (H4) -- +(0, -0.8) node[below, inner sep=1pt] {\large $4$};
    % Top-right V (3)
    \draw[very thick] (H6) -- +(0.65, 0.7) node[above right, inner sep=1pt] {\large $3$};
    \draw[very thick] (H6) -- +(0.85, 0.2);
    % Bottom-right V (3)
    \draw[very thick] (H5) -- +(0.65, -0.7) node[below right, inner sep=1pt] {\large $3$};
    \draw[very thick] (H5) -- +(0.85, -0.2);

    % Center Label
    \node at (0, 0) {\large $AB_1$};
    
    % Left-side labels (single instance of AB_0)
    \node at (-1.6, 0) {\large $AB_0$};
    
    % Right-side labels (single instance of AB_2)
    \node at (2, 0) {\large $AB_2 \cdots$};

    % ==========================================
    % DIAGRAM 2: Regular Pentagon
    % ==========================================
    \coordinate (P1) at ({\Xpent + \Rp*cos(0)}, {\Rp*sin(0)});
    \coordinate (P2) at ({\Xpent + \Rp*cos(72)}, {\Rp*sin(72)});
    \coordinate (P3) at ({\Xpent + \Rp*cos(144)}, {\Rp*sin(144)});
    \coordinate (P4) at ({\Xpent + \Rp*cos(216)}, {\Rp*sin(216)});
    \coordinate (P5) at ({\Xpent + \Rp*cos(288)}, {\Rp*sin(288)});
    
    \draw[very thick] (P1) -- (P2) -- (P3) -- (P4) -- (P5) -- cycle;
    
    % External Legs
    % Top-left V (1)
    \draw[very thick] (P3) -- +(-0.65, 0.7) node[above left, inner sep=1pt] {\large $1$};
    \draw[very thick] (P3) -- +(-0.85, 0.2);
    % Bottom-left V (4)
    \draw[very thick] (P4) -- +(-0.65, -0.7) node[below left, inner sep=1pt] {\large $4$};
    \draw[very thick] (P4) -- +(-0.85, -0.2);
    % Top (2)
    \draw[very thick] (P2) -- +(0.3, 0.7) node[above right, inner sep=1pt] {\large $2$};
    % Right V (3)
    \draw[very thick] (P1) -- +(0.75, 0.3) node[right, inner sep=2pt] {\large $3$};
    \draw[very thick] (P1) -- +(0.75, -0.3);
    % Bottom V (3)
    \draw[very thick] (P5) -- +(0.6, -0.5) ;
    \draw[very thick] (P5) -- +(0.1, -0.75) node[below, inner sep=2pt] {\large $3$};

    % Labels
    \node at (\Xpent, 0) {\large $AB_1$};
    \node at ({\Xpent - 1.8}, 0) {\large $AB_0$};
    \node at ({\Xpent + 1.8}, -0.8) {\large $AB_2 \cdots$};

    % % ==========================================
    % % DIAGRAM 3: Regular Square (Box)
    % % ==========================================
    % \coordinate (B1) at ({\Xbox - \b/2}, {\b/2});
    % \coordinate (B2) at ({\Xbox + \b/2}, {\b/2});
    % \coordinate (B3) at ({\Xbox + \b/2}, {-\b/2});
    % \coordinate (B4) at ({\Xbox - \b/2}, {-\b/2});
    
    % \draw[very thick] (B1) -- (B2) -- (B3) -- (B4) -- cycle;
    
    % % External Legs
    % % Top-left V (1)
    % \draw[very thick] (B1) -- +(-0.65, 0.7) node[above left, inner sep=1pt] {\large $1$};
    % \draw[very thick] (B1) -- +(-0.85, 0.2);
    % % Bottom-left V (4)
    % \draw[very thick] (B4) -- +(-0.65, -0.7) node[below left, inner sep=1pt] {\large $4$};
    % \draw[very thick] (B4) -- +(-0.85, -0.2);
    % % Top-right V (2)
    % \draw[very thick] (B2) -- +(0.65, 0.7) node[above right, inner sep=1pt] {\large $2$};
    % \draw[very thick] (B2) -- +(0.85, 0.2);
    % % Bottom-right V (3)
    % \draw[very thick] (B3) -- +(0.65, -0.7) node[below right, inner sep=1pt] {\large $3$};
    % \draw[very thick] (B3) -- +(0.85, -0.2);

    % % Labels
    % \node at (\Xbox, 0) {\large $AB_1$};
    % \node at ({\Xbox - 1.6}, 0) {\large $AB_0$};
    % \node at ({\Xbox + 1.9}, 0) {\large $AB_2 \cdots$};

  \end{tikzpicture}
\caption{First loop in the Landau diagram being a hexagon (left) or a pentagon (right).}
\label{fig: 1st hexa penta}
\end{figure}
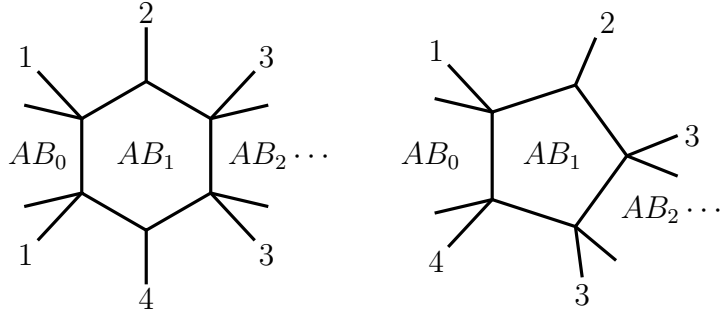
The simplest of these instances is the hexagon case. Indeed, if the line $AB_1$ cuts all external kinematics, as well as its two adjacent loop lines (see Figure~\ref{fig: 1st hexa penta}, left),
\begin{equation}
    \twbr{AB_1 12}=\twbr{AB_1 23} = \twbr{AB_1 34} = \twbr{AB_1 14} = \twbr{AB_0 AB_1} = \twbr{AB_1 AB_2}=0\,,
\end{equation}
then by explicitly solving the first four, the only two possibilities are $AB_1 = (13)$ or $AB_1 = (24)$. This already constrains $z=-1$ just from the cut conditions, which should be satisfied for arbitrary external kinematics and thus doesn't result in an actual singularity.

If, instead, the first loop line is a pentagon (see Figure~\ref{fig: 1st hexa penta}, right), there are several different cases but without loss of generality we can consider
\begin{equation}
    \twbr{AB_1 12}=\twbr{AB_1 23} = \twbr{AB_1 34} = \twbr{AB_0 AB_1} = \twbr{AB_1 AB_2}=0\,,
\end{equation}
since all other possibilities are related by cyclic symmetry. Now, if it turns out that the next loop line satisfies $\twbr{AB_2 23}=0$ (which certainly will be the case if it is a hexagon), one can quickly see that the $AB_1$ pinch condition imposes $\twbr{AB_0 23}=0$, which corresponds to a singularity at $z=0$. At this point, let us mention that other constraints could appear when considering the full Landau diagram, but since these could only lead to higher codimension singularities, we don't have to worry about them.
Having excluded the case when the second loop line $AB_2$ corresponds to a hexagon, we can now enumerate the remaining possible cases:\begin{enumerate}
    \item The second loop line has four-mass box conditions $\twbr{AB_2 12}=\twbr{AB_2 34}=0$. If so, its pinch condition will not have a solution with all Schwinger parameters being non-zero.
    \item The second loop line has three-mass box conditions $\twbr{AB_2 12} = \twbr{AB_2 14}=0$. In that case, its pinch condition will impose $\twbr{AB_1 14}=0$, which means that $AB_1$ cuts all four external lines and we land on a $z=-1$ singularity by the argument above.
    \item The second loop line is a pentagon with $\twbr{AB_2 12}=\twbr{AB_2 34}=\twbr{AB_2 14}=0$. Here, we can simply repeat the same proof as above if the third line is a box, arriving at identical conclusions.
    \item Once there are three or more pentagons in a row, the pinch condition for any intermediate one will not have a solution that keeps all Schwinger proper times non-zero.
\end{enumerate}
The only special case arises at two loops where both $AB_1$ and $AB_2$ are pentagons, which we have already discussed explicitly in Section~\ref{sec: examples}. Altogether, we have now established that any Landau diagram where the loop line $AB_1$ is associated with a hexagon or pentagon can only lead to singularities at $z\in\{-1,0,\infty\}$.

\subsection{Four-mass box as the first loop line}\label{sec: 4mass box}

\begin{figure}
\centering
  \begin{tikzpicture}[baseline=(current bounding box.center), scale=0.9]
    
    % Define a smaller shared side length for compactness
    \def\a{1.1}
    \def\b{1.6}
    \def\Rp{1.1} % Radius for regular pentagon: a / (2 * sin(36))
    
    % Spread out the centers to ensure absolutely no overlap
    \def\Xpent{6} % Center X of pentagon
    \def\Xbox{11.5} % Center X of box

    \coordinate (B1) at ({\Xbox - \b/2}, {\b/2});
    \coordinate (B2) at ({\Xbox + \b/2}, {\b/2});
    \coordinate (B3) at ({\Xbox + \b/2}, {-\b/2});
    \coordinate (B4) at ({\Xbox - \b/2}, {-\b/2});
    
    \draw[very thick] (B1) -- (B2) -- (B3) -- (B4) -- cycle;
    
    % External Legs
    % Top-left V (1)
    \draw[very thick] (B1) -- +(-0.65, 0.7) node[above left, inner sep=1pt] {\large $1$};
    \draw[very thick] (B1) -- +(-0.85, 0.2);
    % Bottom-left V (4)
    \draw[very thick] (B4) -- +(-0.65, -0.7) node[below left, inner sep=1pt] {\large $4$};
    \draw[very thick] (B4) -- +(-0.85, -0.2);
    % Top-right V (2)
    \draw[very thick] (B2) -- +(0.65, 0.7) node[above right, inner sep=1pt] {\large $2$};
    \draw[very thick] (B2) -- +(0.85, 0.2);
    % Bottom-right V (3)
    \draw[very thick] (B3) -- +(0.65, -0.7) node[below right, inner sep=1pt] {\large $3$};
    \draw[very thick] (B3) -- +(0.85, -0.2);

    % Labels
    \node at (\Xbox, 0) {\large $AB_1$};
    \node at ({\Xbox - 1.6}, 0) {\large $AB_0$};
    \node at ({\Xbox + 1.9}, 0) {\large $AB_2 \cdots$};

  \end{tikzpicture}
\caption{First loop in the Landau diagram is a four-mass box.}
\label{fig: 1st 4mass box}
\end{figure}
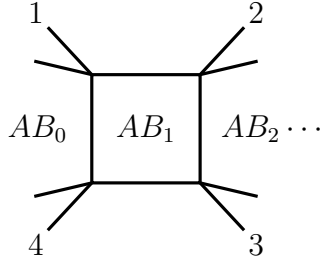

We move on to the set of Landau diagrams for which $AB_1$ satisfies four-mass box conditions (see Figure~\ref{fig: 1st 4mass box}, right), such as
\begin{equation}
    \twbr{AB_1 12} = \twbr{AB_1 34} = \twbr{AB_0 AB_1} = \twbr{AB_1 AB_2} =0\,.
\end{equation}
(The only other possibility would be to cycle $Z_i \to Z_{i+1}$). Generically, we can write the pinch condition for $AB_1$ as the following matrix equation:
\begin{equation}
    \begin{pmatrix}
        0&\twbr{1234}&\twbr{AB_{0}12}&\twbr{AB_2 12}\\
        \twbr{1234}&0&\twbr{AB_034}&\twbr{AB_2 34}\\
        \twbr{AB_{0}12}&\twbr{AB_{0}34}&0&\twbr{AB_{0}AB_{2}}\\
        \twbr{AB_2 12}&\twbr{AB_2 34}&\twbr{AB_{0}AB_{2}}&0
    \end{pmatrix}
    \vec{\alpha} =0\,.
\end{equation}
There are now three distinct possibilities, depending on the second loop line. Firstly, if $AB_2$ shares two of the external cut conditions with $AB_1$ (which can happen whether the former is a hexagon, pentagon or box), then the pinch condition for the latter doesn't have a solution. Secondly, if $AB_2$ only shares one of them, e.g. $\twbr{AB_2 12}=0$, the pinch condition for $AB_1$ will constrain $\twbr{AB_0 12}=0$, which corresponds to a singularity at $z=0$. Finally, the most subtle situation arises if $AB_2$ is another four-mass box satisfying
\begin{equation}
    \twbr{AB_2 23} = \twbr{AB_2 14} = \twbr{AB_1 AB_2} = \twbr{AB_2 AB_3} =0\,.
\end{equation}
To analyze this case let us explicitly parameterize the loop lines. The generic solution to the subset of cut equations involving only the external kinematics $Z_i$ is
\begin{equation}
\begin{aligned}
    AB_{1} &= \left(\alpha_1 \,1 + (1-\alpha_1)2,\, \beta_1 \,3 + (1-\beta_1)4\right),\\
    AB_{2} &= \left( \gamma_2 \, 2 + (1-\gamma_2)3,\, \delta_2 4 - (1-\delta_2)1 \right),
\end{aligned}
\end{equation}
with $0\leq\alpha_1,\,\beta_1,\,\gamma_2,\,\delta_2\leq1$ in order to satisfy~(\ref{eq: amplituhedron ext positivity}). The cut condition between $AB_1$ and $AB_2$ then requires that
\begin{equation}
\begin{aligned}
    \twbr{AB_{1}AB_{2}} &= -\alpha_1\beta_1\gamma_2\delta_2 - (1-\alpha_1)(1-\beta_1)(1-\gamma_2)(1-\delta_2) \\&= -\twbr{AB_{1}24}\twbr{AB_{2}13} - \twbr{AB_{1}13}\twbr{AB_{2}24}
\end{aligned}
\end{equation}
must vanish, but the conditions~(\ref{eq: amplituhedron ext positivity}) enforce each of the two terms on the second line to be strictly non-positive. Since we are only imposing the cut conditions at this stage, we must remain inside the (closure of) the geometry, and thus we need each term to vanish separately. In other words:
\begin{equation}\label{eq: 4mass box int cut}
    \twbr{AB_{1}24}\twbr{AB_{2}13} = \twbr{AB_{1}13}\twbr{AB_{2}24}=0\,.
\end{equation}
Now note that if $\twbr{AB_1 13}=\twbr{AB_1 24}=0$, then it fixes either $AB_1 = (23)$ or $AB_1 = (14)$. By the cut condition with the frozen loop line, this implies $\twbr{AB_0 23}=0$ or $\twbr{AB_0 14}=0$, which indicates a singularity at $z=\infty$. On the other hand, if $\twbr{AB_2 13}\twbr{AB_2 24}=0$ we can use the Schouten identity
\begin{equation}
    \twbr{AB_2 12}\twbr{AB_2 34} + \twbr{AB_2 23}\twbr{AB_2 14} = \twbr{AB_2 13}\twbr{AB_2 24}
\end{equation}
to see that $\twbr{AB_2 12}\twbr{AB_2 34}=0$. The pinch condition for $AB_1$ then imposes the constraint $\twbr{AB_0 12}\twbr{AB_0 34}=0$, which gives a singularity at $z=0$. Altogether, we have established that the statement made at the end of Section~\ref{sec: hexagon or pentagon} holds true for four-mass boxes as well.

\subsection{Three-mass box as the first loop line}\label{sec: 3mass box}

The only Landau diagrams left to consider are those for which the loop line $AB_1$ is associated with a box with three-mass cut conditions; for example
\begin{gather}
    \twbr{AB_1 12} = \twbr{AB_1 23} = \twbr{AB_0 AB_1} = \twbr{AB_1 AB_2}=0\,.\\
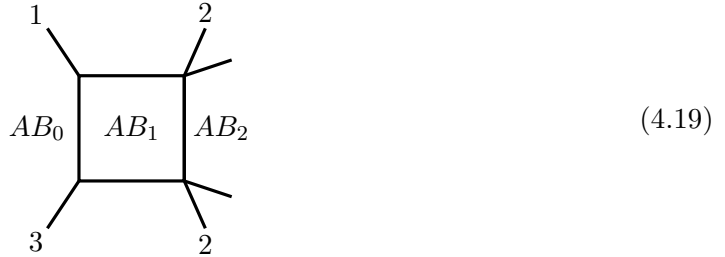
    \begin{tikzpicture}[baseline=(current bounding box.center), scale=0.7]
    % --- Left Block (Boxes AB_1, AB_2) ---
    \draw[very thick] (0,0) rectangle (2,2);
    \draw[very thick] (2,0) -- (2,2);
    \node at (-0.8, 1) {$AB_0$};
    \node at (1, 1) {$AB_1$};
    \node at (2.7, 1) {$AB_2$};
    % External legs on left of AB_1
    \draw[very thick] (0,2) -- +(-0.6, 0.9) node[above left, inner sep=1pt] {$1$};
    \draw[very thick] (0,0) -- +(-0.6, -0.9) node[below left, inner sep=1pt] {$3$};
    % Shared vertical legs (V-shapes) between AB_1 and AB_2
    \draw[very thick] (2,2) -- +(0.4, 0.9) node[above, inner sep=2pt] {$2$};
    \draw[very thick] (2,2) -- +(0.9, 0.3) node[above, inner sep=2pt] {};
    \draw[very thick] (2,0) -- +(0.4, -0.9) node[below, inner sep=2pt] {$2$};
    \draw[very thick] (2,0) -- +(0.9, -0.3) node[below, inner sep=2pt] {};
    \end{tikzpicture}
\end{gather}
The pinch condition for $AB_1$ is
\begin{equation}\label{eq: 3mass box int cut}
\begin{pmatrix}
    0 & 0 & \twbr{AB_0 12} & \twbr{AB_2 12} \\
    0 & 0 & \twbr{AB_0 23} & \twbr{AB_2 23} \\
    \twbr{AB_0 12} & \twbr{AB_0 23} & 0 & \twbr{AB_0 AB_2}\\
    \twbr{AB_2 12} & \twbr{AB_2 23} & \twbr{AB_0 AB_2} & 0
\end{pmatrix} \vec{\alpha} = 0\,.
\end{equation}
We can easily check that if the second loop line has an overlapping cut condition $\twbr{AB_2 12}=0$ or $\twbr{AB_2 23} = 0$ (regardless of its overall topology), then the pinch condition for the leading singularity implies that $\twbr{AB_0 12} = \twbr{AB_0 23} = 0$. This is a codimension-2 singularity we don't need to worry about. The only remaining possibility is that the second loop line is another three-mass box with cut conditions that don't overlap with those of $AB_1$, i.e.
\begin{gather}
\begin{aligned}
    &\twbr{AB_1 12} = \twbr{AB_1 23} = \twbr{AB_0 AB_1} = \twbr{AB_1 AB_2} = 0\,,\\
    &\twbr{AB_2 34} = \twbr{AB_2 14} = \twbr{AB_2 AB_3} = 0\,,\\
\end{aligned} \\
    \begin{tikzpicture}[baseline=(current bounding box.center), scale=0.7]
    % --- Left Block (Boxes AB_1, AB_2) ---
    \draw[very thick] (0,0) rectangle (4,2);
    \draw[very thick] (2,0) -- (2,2);
    \node at (-0.8, 1) {$AB_0$};
    \node at (1, 1) {$AB_1$};
    \node at (3, 1) {$AB_2$};
    \node at (4.8, 1) {$AB_3$};
    % External legs on left of AB_1
    \draw[very thick] (0,2) -- +(-0.5, 0.9) node[above left, inner sep=1pt] {$1$};
    \draw[very thick] (0,0) -- +(-0.5, -0.9) node[below left, inner sep=1pt] {$2$};
    % Shared vertical legs (V-shapes) between AB_1 and AB_2
    \draw[very thick] (2,2) -- +(-0.3, 0.9) node[above, inner sep=2pt] {$2$};
    \draw[very thick] (2,2) -- +(0.3, 0.9) node[above, inner sep=2pt] {$3$};
    \draw[very thick] (2,0) -- +(-0.3, -0.9) node[below, inner sep=2pt] {$3$};
    \draw[very thick] (2,0) -- +(0.3, -0.9) node[below, inner sep=2pt] {$4$};
    % External legs on right of AB_2
    \draw[very thick] (4,2) -- +(0.4, 0.9) node[above right, inner sep=1pt] {$4$};
    \draw[very thick] (4,0) -- +(0.4, -0.9) node[below right, inner sep=1pt] {$1$};
    \end{tikzpicture}
\end{gather}
The cut conditions for $AB_1$ are solved by
\begin{equation}
\begin{aligned}
    AB_1 &= (2 AB_{0})\cap (2 AB_{2}) ~ {\rm or}\\
    AB_1 &= \big( (123)\cap AB_{0},(123)\cap AB_{2} \big),
\end{aligned}
\end{equation}
and its pinch condition \eqref{eq: 3mass box int cut} results in the following constraints for $AB_2$:
\begin{equation}\label{eq: pinch condition 3mass box chain}
    \twbr{AB_{0}AB_{2}}=0\,,\quad \frac{\twbr{AB_{0}12}}{\twbr{AB_{0}23}}=\frac{\twbr{AB_{2}12}}{\twbr{AB_{2}23}}\,.
\end{equation}
The second equation in \eqref{eq: pinch condition 3mass box chain} can be rewritten using \eqref{eq:plane int id} in the following way:
\begin{equation}
    \twbr{123\, AB_{0}\cap AB_{2}}=0\,,
\end{equation}
where we have used that $AB_{0}$ and $AB_{2}$ have to intersect thanks to~(\ref{eq: pinch condition 3mass box chain}). In other words, the intersection between these two loop lines has to lie in the (123) plane. Going back to our solutions for $AB_1$, we see that the second one is degenerate since both points are the same, while the first one can be written as
\begin{equation}
\label{eq:cond}
    AB_1 = \big(2,AB_{0}\cap AB_{2})\,.
\end{equation}
Thus, $AB_1$ both passes through the point $Z_2$ and lies in the (123) plane. In exactly the same vein, the second loop line $AB_2$ passes through $Z_4$ and lies in the (341) plane (assuming that the cut conditions for $AB_3$ have no overlap with those of $AB_2$; if they do, then the pinch condition for $AB_2$ implies that  $\twbr{AB_1 34} = \twbr{AB_1 14} = 0$, which implies a singularity at $z=-1$).   Now it follows from~(\ref{eq:cond}) that $AB_0$, $AB_1$ and $AB_2$ all intersect at a common point, which must therefore be on the line $(13)$. Therefore $\langle AB_0 13\rangle = 0$, and we see that there is a singularity at $z=-1$.

Before concluding let us return to the subtlety mentioned in Section~\ref{sec: landau} and discussed previously at the end of Section~\ref{sec: one-loop analysis}. Suppose there is a Landau diagram whose associated equations have a codimension-1 solution with some $\alpha_e = 0$ and the corresponding solution of the associated subtopology is generic (i.e., codimension-0).  From our analysis this can only happen if all other loops in the subtopology are bubbles.  But we have seen that bubbles can be trivially eliminated, so it is sufficient to check that nothing is overlooked at one loop, which we have already done in Section~\ref{sec: one-loop analysis}.

This concludes the proof that the $L$-loop one-cycle negative geometry, given by integrating~(\ref{eq:onecycle}), can have branch point singularities only at $z \in\{ -1, 0, \infty \}$.

\section{Outlook}\label{sec: outlook}

In this paper we have used the tools of geometric Landau analysis~\cite{Dennen:2016mdk} to classify the singularities of all four-point, one-cycle negative geometries. By examining general features of the associated Landau diagrams, we have proven that physical singularities can occur only at three values $z\in\{-1,0,\infty\}$ of the kinematic variable~(\ref{eq: z variable}), to all orders in the loop expansion. The simplicity of this result leads to several open questions.

Firstly, even for functions that evaluate to multiple polylogarithms, it is well-known that knowledge of the location of physical singularities is not enough to completely determine the symbol alphabet.  For example, the putative letters $\frac{1 + \sqrt{-z}}{1 - \sqrt{-z}}$ and $\frac{1+\sqrt{1+z}}{1-\sqrt{1+z}}$ have branch-point singularities at $z=-1$ and $0$, respectively. Indeed, upcoming work~\cite{Dixon:2026} shows that the former is a symbol letter of individual negative geometry graphs (though not of the types of graphs we considered here) that cancels out of the sum. It is natural to ask whether there is a simple set of physically motivated assumptions that might inspire a guess for the symbol letters that appear in one-cycle geometries at arbitrary loop order.

At two and three loops, the explicit known results for the (integrated) four-point negative geometries (i.e., logarithms of four-point amplitudes~\cite{Anastasiou:2003kj,Bern:2005iz}) indicate that they are expressible as linear combinations of harmonic polylogarithm functions (HPLs) with singularities only at $z\in\{-1,0,\infty\}$. Our results hint that this may hold for higher-loop geometries as well. This is particularly interesting because \cite{Dixon:2026} argues that one can resum the series of diagrams with a specific one-cycle graph (e.g. a pentagon as shown below) attached to an arbitrary tree, of the type
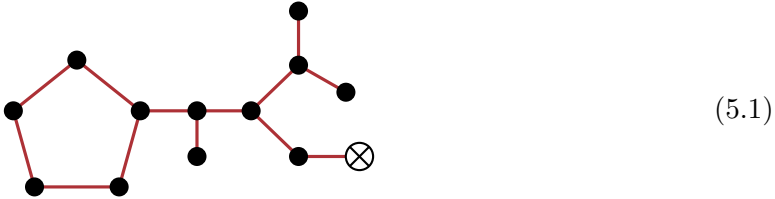
\begin{equation}\label{eq: cycle_series}
\begin{tikzpicture}[baseline={([yshift=-1.5ex]current bounding box.center)},
    thick, scale=0.9,
    cross/.style={path picture={
      \draw[black] (path picture bounding box.south east) -- (path picture bounding box.north west) (path picture bounding box.south west) -- (path picture bounding box.north east);
    }}]
    % Pentagon coordinates
    \coordinate (AB) at (0,0);
    \coordinate (CD) at (-0.3125, 1.125);
    \coordinate (EF) at (0.625, 1.875);
    \coordinate (GH) at (1.5625, 1.125);
    \coordinate (IK) at (1.25,0);
    
    % New Tree coordinates (spaced out horizontally)
    \coordinate (N0) at (2.4, 1.125); % First node in horizontal chain
    \coordinate (N1) at (3.2, 1.125); % Second node (Branch point)
    \coordinate (N7) at (2.4, 0.45);
    \coordinate (N2) at (3.9, 1.8);   % Upper sub-branch
    \coordinate (N3) at (3.9, 0.45);  % Lower sub-branch
    \coordinate (N4) at (3.9, 2.6);   % Terminal node 1 (Top)
    \coordinate (N5) at (4.6, 1.4);   % Terminal node 2 (Middle)
    \coordinate (N6) at (4.8, 0.45);  % Terminal node 3 (Bottom right)
    \coordinate (N6b) at (4.6, 0.45);

    % Draw the pentagon
    \draw[very thick, Maroon] (AB) -- (CD) -- (EF) -- (GH) -- (IK) -- (AB);
    
    % Draw the tree edges
    \draw[very thick, Maroon] (GH) -- (N0);
    \draw[very thick, Maroon] (N0) -- (N1);
    \draw[very thick, Maroon] (N1) -- (N2);
    \draw[very thick, Maroon] (N1) -- (N3);
    \draw[very thick, Maroon] (N2) -- (N4);
    \draw[very thick, Maroon] (N2) -- (N5);
    \draw[very thick, Maroon] (N3) -- (N6b);
    \draw[very thick, Maroon] (N0) -- (N7);

    % Draw the pentagon vertices
    \fill (AB) circle (4pt);
    \fill (CD) circle (4pt);
    \fill (EF) circle (4pt);
    \fill (GH) circle (4pt);
    \fill (IK) circle (4pt);
    
    % Draw the internal tree vertices and top terminal nodes
    \fill (N0) circle (4pt);
    \fill (N1) circle (4pt);
    \fill (N2) circle (4pt);
    \fill (N3) circle (4pt);
    \fill (N4) circle (4pt);
    \fill (N5) circle (4pt);
    \fill (N7) circle (4pt);
    
    % Draw the lowest right terminal vertex using the cross style
    \node[draw,circle,cross,minimum width=8pt] at (N6) {};
  \end{tikzpicture}
\end{equation}
if the former consists only of HPLs. If one could obtain the precise combination of HPLs that make up the one-cycle graphs we discuss, it would be possible to resum the result to all loop orders. This would calculate the next non-perturbative contribution to the cusp anomalous dimension in the expansion over cycles. Further, upcoming work \cite{Dixon:2026} indicates that the different series composed of HPL functions contribute as $g^{2L-1}$ at strong coupling, where $L$ is the number of loops in the original one-cycle geometry. As a result, it is highly likely that a sensible result for the resummation can only be obtained after all possible one-cycle geometries are taken into account, rather than individual series such as \eqref{eq: cycle_series}.

Finally, a couple of natural extensions to this work would be to analyze the singularity structure of higher-cycle and higher-point negative geometries. Higher-point leading singularities have been analyzed in \cite{Brown:2025plq}. These should also appear as solutions to the Landau equations, and it would be interesting to study their interplay from this point of view. We leave these questions for future work.

\acknowledgments

We are grateful to Nima Arkani-Hamed, Lance Dixon, Song He, Elia Mazzucchelli, Jaroslav Trnka, Cristian Vergu and Qinglin Yang for useful discussions, and to Akshay Yelleshpur Srikant for collaboration in the early stages of this work. This work was supported in part by the US Department of Energy under contract DE-SC0010010 Task F and by Simons Investigator Award \#376208 (SP, AV).

\bibliographystyle{JHEP}
\bibliography{bibliography.bib}

\end{document}